\documentstyle[12pt]{article}
\def\be{\begin{equation}}
\def\ee{\end{equation}}
\def\bc{\begin{center}}
\def\ec{\end{center}}
\def\bea{\begin{eqnarray}}
\def\eea{\end{eqnarray}}
\def\nn{\nonumber}
\def\dd{\displaystyle}

\def\mev{{\rm \; MeV}}
\def\gev{{\rm \; GeV}}

\catcode`@=11
\def\marginnote#1{}
\newcount\hour
\newcount\minute
\newtoks\amorpm
\hour=\time\divide\hour by60
\minute=\time{\multiply\hour by60 \global\advance\minute by-\hour}
\edef\standardtime{{\ifnum\hour<12 \global\amorpm={am}%
        \else\global\amorpm={pm}\advance\hour by-12 \fi
        \ifnum\hour=0 \hour=12 \fi
        \number\hour:\ifnum\minute<10 0\fi\number\minute\the\amorpm}}
\edef\militarytime{\number\hour:\ifnum\minute<10 0\fi\number\minute}
\def\draftlabel#1{{\@bsphack\if@filesw {\let\thepage\relax
   \xdef\@gtempa{\write\@auxout{\string
      \newlabel{#1}{{\@currentlabel}{\thepage}}}}}\@gtempa
   \if@nobreak \ifvmode\nobreak\fi\fi\fi\@esphack}
        \gdef\@eqnlabel{#1}}
\def\@eqnlabel{}
\def\@vacuum{}
\def\draftmarginnote#1{\marginpar{\raggedright\scriptsize\tt#1}}
\def\draft{\oddsidemargin 0.0truein
        \def\@oddfoot{\sl preliminary draft \hfil
        \rm\thepage\hfil\sl\today\quad\militarytime}
        \let\@evenfoot\@oddfoot \overfullrule 3pt
        \let\label=\draftlabel
        \let\marginnote=\draftmarginnote
   \def\@eqnnum{(\theequation)\rlap{\kern\marginparsep\tt\@eqnlabel}%
\global\let\@eqnlabel\@vacuum}  }
\catcode`@=12
\begin{document}
\begin{titlepage}
\vspace*{4.cm}
\begin{center}
{\Large\bf
TWO-LOOP CORRECTIONS\\
\vskip .1cm
FOR\\

\vskip .3cm
ELECTROWEAK PROCESSES}
\footnote{Presented by F. Feruglio}
\footnote{Work supported in part by the European Union under
contract No.~CHRX-CT92-0004.}
\end{center}
\vskip 1.0cm
\begin{center}
{\large Giuseppe Degrassi, Ferruccio Feruglio, Alessandro Vicini}\\
\vskip .1cm
Dipartimento di Fisica, Universit\`a di Padova, I-35131 Padua, Italy
\\
\vskip .2cm
and
\vskip .2cm
{\large Sergio Fanchiotti, Paolo Gambino}
\\
\vskip .1cm
Dept. of Physics, New York University,
\\
4 Washington Place, New York, NY 10003 U.S.A.
\\
\vskip 5.cm
\end{center}
\vskip 0.5cm
\begin{abstract}
\noindent
Theoretical uncertainties affecting electroweak observables are reviewed
and the relevance of two-loop electroweak radiative corrections for the
precision tests of the Standard Model is discussed.
\end{abstract}
\vfill
\end{titlepage}
\setcounter{footnote}{0}
\vskip2truecm
%
The increasing accumulation of electroweak data over the last years
have brought to full maturity the program of precision tests of the
Standard Model (SM).
One of the most remarkable outcome of such a program is
the determination of the top quark mass from a fit of
the available data. By considering the LEP data alone,
one finds \cite{cal}
\footnote{
The free parameters in the fit are $m_t$
and $\alpha_s$. The Higgs mass $m_H$ is fixed at $300$ GeV. The last error
in eq. (\ref{topfit})
shows the effect of varying $m_H$ from $60$ GeV to 1 TeV. The same fit
with the inclusion of all low-energy and SLD data gives
$m_t=179\pm9^{+17}_{-19}$.
}:
\be
m_t=176\pm10^{+17}_{-19}\gev
\label{topfit}
\ee
with a $\chi^2$ per degree of freedom of 8.8/9.
It has been known since a long time that the potentially largest
electroweak radiative corrections are those related to $m_t$ \cite{vel}.
In the absence of a direct evidence
for the top quark, this property of the SM has been exploited to
indirectly bound $m_t$. The absolute accuracy reached in
constraining the top mass is now impressive and
the good quality of the fit provides a strong
consistency check of the SM beyond the tree level approximation.
In addition, there is a striking agreement between the above quoted
value of $m_t$ and the recent results of a direct search of the top
quark at the Tevatron collider \cite{cdf,d0,top}:
\bea
m_t&=&176\pm 8({\rm stat})\pm 10({\rm sys})\gev~~~~~~~{\rm CDF}\nn\\
m_t&=&199^{+19}_{-21}({\rm stat})\pm 22({\rm sys})\gev~~~~~~~{\rm D0}
\label{mtop}
\eea
Such an agreement puts on a very sound basis our understanding
and interpretation of the electroweak data within the framework of the SM
as a renormalizable quantum field theory and places severe constraints
on possible extensions of the model.

A closer look to the data reveals that the achieved experimental
precision went by far beyond the initial expectations \cite{phy}. As an
example,
consider the most recent LEP data for the width of $Z$
into charged leptons $\Gamma_l$, the effective Weinberg angle in the
leptonic sector $\theta^l_{eff}$ and the $W$ mass $M_W$ \cite{cal}:
\bea
\Gamma_l&=&83.94\pm 0.13\mev\nn\\
\sin^2\theta_{eff}^l&=&0.2318\pm0.0004\nn\\
M_W&=&80.26\pm0.16\gev~~~~~
\label{data}
\eea
The relative accuracy in these measurements has reached the level of
$1-2$ per mill.
It is conceivable that the experimental precision may further
improve in the future. The completion of the LEP I phase
could lead to an overall amount of $2\cdot 10^7$ $Z$, whereas
the SLAC program aims for a total of $5\cdot 10^5$ $Z$ by the end
of 1998. This could reduce the error on $\sin^2\theta_{eff}^l$
to $0.0002$. On the other hand, the measurement of $M_W$ with
an error of $50~{\rm MeV}$ is one of the main tasks of the LEP II program.

Given these remarkable achievements and future perspectives,
a natural question arises. Does
the theoretical error affecting the predictions of the SM
match the present/expected experimental precision?
This question has been recently addressed and discussed by
a Working Group on Precision Calculation organized in 1994
at CERN. Main purpose of this Working Group has been
to update the predictions of the SM concerning the observables
of interest at LEP and to estimate the theoretical uncertainties
of these predictions. Here we will summarize very concisely
some of the results of the Working Group.
For a complete information, we address the reader
to the full Report \cite{rep}.

The possible sources of theoretical uncertainties can be
divided into two general classes: parametric uncertainties
and intrinsic uncertainties.
The former are related to the fact that, within the SM, each quantity
of interest is a function of a set of input parameters, which are known
with a finite experimental precision. Any variation of the input parameters
within the experimentally allowed range gives rise to an
uncertainty on the observable considered. On the other hand,
the intrinsic uncertainties have to do with
the perturbative treatment of the quantum corrections: scheme dependence,
ignorance of higher orders in the perturbative expansion and so on.

Among the input parameters of the SM, the fine structure constant $\alpha$,
the Fermi constant $G_\mu$ and the $Z$ mass $M_Z$ enter the expressions
for the electroweak observables already at tree level, and therefore
are immediately involved in the discussion of the parametric
uncertainties. At one-loop order the dependence is extended to the
other parameters, such as the strong coupling $\alpha_s$ and all
the masses of the SM particles.
By inspecting the experimental accuracy of the values for the three
basic parameters
\bea
\alpha^{-1}&=&137.0359895(61)\nn\\
G_{\mu}&=&1.16639(2)\cdot 10^{-5}\gev^{-2}\nn\\
M_Z&=&91.1887\pm0.0022\gev~~~,
\label{input}
\eea
one would conclude that the parametric uncertainties they induce
are, for each quantity of interest,
well below the present and future experimental sensitivity.

However, it is well known that, after the inclusion of quantum corrections,
the observables at the $Z$ resonance exhibit an explicit dependence
on the fine structure constant $\alpha$ only through the combination
${\bar \alpha}$, which accounts for the running of $\alpha$ from $q^2=0$
to $q^2=M_Z^2$. The effective constant ${\bar \alpha}$
is related to $\alpha$ by the relation:
\be
{\bar \alpha}=\frac{\alpha}{1-\Delta\alpha}
\label{alpb}
\ee
The correction $\Delta\alpha$ contains the fermionic contribution
to the photon vacuum polarization, usually split into leptonic
and hadronic parts:
\be
\Delta\alpha=\Delta\alpha_l+\Delta\alpha_h
\label{dalp}
\ee
The leptonic part $\Delta\alpha_l$ is computed perturbatively and
known with very high precision \cite{rep,kni}.
On the contrary the hadronic part, involving exchange of gluons
at low momenta, cannot be reliably computed in perturbation theory.
An estimate of $\Delta\alpha_h$ is then obtained by relating it
to the cross-section for $e^+e^-$ into hadrons via the dispersion
relation:
\bea
\Delta\alpha_h&=&\frac{\alpha M_Z^2}{3\pi} \int_{4 m_\pi^2}^\infty
ds \frac{R(s)}{s(M_Z^2-s)}\nn\\
R(s)&=&\frac{\sigma(e^+e^-\to\gamma\to {\rm hadrons})}
{\dd\left(\frac{4\pi \alpha^2}{3 s}\right)}
\label{dalph}
\eea
The low-energy part of the cross-section
$\sigma(e^+e^-\to\gamma\to{\rm hadrons})$ is directly taken from
the existing data whose experimental error is
propagated to $\Delta\alpha_h$. The estimate of $\Delta\alpha_h$
which has been used up to now in the electroweak libraries
is the one of ref. \cite{jeg1}:
\be
\Delta\alpha_h=0.0282\pm0.0009
\label{dalphh}
\ee
which leads to ${\bar\alpha}^{-1}=128.87\pm 0.12$. The sensitivity
of the electroweak data to this error can be illustrated by
considering the uncertainties induced on $\Gamma_l$, $\sin^2\theta^l_{eff}$
and $M_W$. One obtains:
\bea
\delta\Gamma_l&=&0.014\mev\nn\\
\delta\sin^2\theta^l_{eff}&=&0.0003\nn\\
\delta M_W&=&16\mev
\label{shift}
\eea
Whereas the uncertainties on $\Gamma_l$ and on $M_W$ are negligible,
the error on $\sin^2\theta^l_{eff}$ is already comparable to the present
experimental accuracy.
Recent re-evaluations of $\Delta\alpha_h$ \cite{dal}, reported at this meeting
by Burkhardt, reduce somewhat the above quoted uncertainty,
but not at the level required by the future experimental sensitivity
on $\sin^2\theta^l_{eff}$.
We refer to ref. \cite{bur} for a detailed discussion
of this point.

Coming to the intrinsic uncertainties, as mentioned before,
they are related to the use of perturbation theory in the
evaluation of the quantum corrections. In particular, since
the existing electroweak libraries
work at least at the one-loop approximation, the intrinsic uncertainties
are presently due to the incomplete knowledge of two or higher-loop terms.

One source of error is the dependence on the chosen renormalization
scheme. Different renormalization schemes are available for the
computation of loop corrections. By working at a fixed order of
perturbation theory, e.g. in the one-loop
approximation, all the schemes provide consistent results.
This
implies that the differences among the results obtained in the
various schemes are of higher, in this case at least two-loop,
order. These differences, although formally negligible at one-loop level,
can be in practice numerically important.

Other intrinsic uncertainties are related to the splitting
of a given correction into a leading part plus a remainder.
To exemplify this, consider the usual parametrization of the
$Z$ width $\Gamma_f$ into a fermion pair $f {\bar f}$:
\be
\Gamma_f=4 N_c^f \frac{G_\mu M_Z^3}{24 \pi\sqrt{2}} \left[
         (g_V^f)^2+(g_A^f)^2\right]
\label{width}
\ee
where final QED and QCD corrections have been removed.
The effective vector and axial-vector couplings $g_V^f$ and $g_A^f$
are in turn expressed in terms of the parameters $k_f$ and $\rho_f$
by the relations:
\be
\frac{g_V^f}{g_A^f}=1-4|Q^f_{em}| s_W^2 k_f,~~~~~~~~~~
4(g_A^f)^2=\rho_f
\label{gagv}
\ee
where $s_W^2=1-M_W^2/M_Z^2$. In this way $k_f$ defines an effective
weak mixing angle for the flavour $f$, whereas $\rho_f$ measures
the overall strength of the neutral current interaction in the
$f {\bar f}$ channel.

The $\rho_f$ parameter can be further decomposed into two
contributions:
\be
\rho_f=\frac{1}{(1-\Delta\rho)}+\Delta\rho_{f,rem}
\label{split}
\ee
The first term in eq. (\ref{split}) is the so-called leading
contribution. The correction $\Delta\rho$ involves all gauge-invariant
fermionic contributions to $W$ and $Z$ self-energy diagrams.
This correction is universal, that is it does not depend on the
flavour $f$, and contains the potentially largest effect, related
to positive powers of the top mass:
\be
\Delta\rho=N_c x_t \left[1+x_t \Delta\rho^{(2)}(\frac{m_H^2}{m_t^2})+
                         c_1 \frac{\alpha_s(M_Z)}{\pi}+
                         c_2 (\frac{\alpha_s(M_Z)}{\pi})^2 +...\right]
\label{drho}
\ee
where $N_c=3$, $x_t=\dd\frac{G_\mu m_t^2}{8 \pi^2\sqrt{2}}$. After extraction
of the common factor $N_c x_t$, the unity represents the well known one-loop
contribution \cite{vel} which, for values of $m_t$ in the range given in eqs.
(\ref{topfit},\ref{mtop}) is about 0.01;
$\Delta\rho^{(2)}$ is the leading electroweak
two-loop result in units of $N_c x_t^2$ \cite{bar}
and the remaining terms are the known
mixed electroweak-strong corrections \cite{ver,adv,che}.
These last corrections are relatively large. The $\alpha_s$ contribution
\cite{ver} is about ten per cent of the one-loop correction
$N_c x_t$. Very recently two groups
\cite{adv,che} have computed independently the $\alpha_s^2$ term
obtaining $c_2=-\pi^2~(2.56-0.18 n_f)$, $n_f$ being the total number of
quarks. For $n_f=6$ one finds $c_2=-14.6$ and the $\alpha_s^2$ contribution
is about 20\% of the $\alpha_s$ term. It is reasonable
to think that the theoretical errors coming from neglecting higher-orders in
the $\alpha_s$ series are below the experimental sensitivity \cite{sir}.
Finally, notice that in eq. (\ref{split}) the Dyson series obtained by
iterating the correction $\Delta\rho$ has been resummed.

On the contrary, the remainder $\Delta\rho_{f,rem}$ is flavour dependent
and of order $G_\mu M_Z^2$. Moreover, since it involves the
combination of vertex and self-energy bosonic corrections, which are not
separately gauge invariant, there is no simple way to resum
iterations of these contributions.
To simulate the effect of higher order contributions to $\rho_f$,
not explicitly accounted for by the equation (\ref{split}),
some of the existing libraries have considered different "options".
These options consist in alternative expressions for $\rho_f$
such as:
\be
(\rho_f)'=\frac{1}{(1-\Delta\rho-\Delta\rho_{f,rem})},~~~~~~~~~~
(\rho_f)''=\frac{1}{(1-\Delta\rho)}~
\left[1+\frac{\Delta\rho_{f,rem}}{(1-\Delta\rho)}\right]
\label{options}
\ee
The expressions in eqs. (\ref{split},\ref{options}) are
equivalent at one-loop level. The differences are of order
$\Delta\rho\cdot\Delta\rho_{f,rem}\simeq O(G_\mu^2 m_t^2 M_Z^2)$
and can be fixed only by an explicit two-loop computation.
In the absence of such a computation one may try to estimate
the correspondent theoretical uncertainty by running the code
with different options and by looking at the range of values
obtained for a given observable.

Similarly, further theoretical uncertainties arise from the ignorance
of two-loop terms in the electroweak expansion
concerning the parameters $k_f$ and $\Delta r$. Presently
unknown two and higher-loop
contributions would also remove ambiguities related to the
factorization of QCD and electroweak corrections and to the linearization
of expressions containing
squares of tree level plus one-loop contributions. To size the
effects on electroweak observables, each code
has foreseen different options, similar to those described above
for the $\rho_f$ parameter. By comparing the results obtained
by choosing different options one has an indication about the
importance of the higher-order corrections which have not yet been
explicitly computed.

As an example, we consider below the values of $\sin^2\theta^l_{eff}$
computed in the SM by the existing electroweak libraries, at the
reference point $m_t=175\gev$, $m_H=300\gev$, $\alpha_s(M_Z)=0.125$.
\be
\sin^2\theta^l_{eff}=
\left\{
\begin{array}{cc}
0.23197^{+0.00004}_{-0.00007}&{\rm BHM} \cite{bhm}\\

0.23200^{+0.00008}_{-0.00008}&{\rm LEPTOP} \cite{lep}\\

0.23200^{+0.00004}_{-0.00004}&{\rm TOPAZ0} \cite{paz}\\

0.23194^{+0.00003}_{-0.00007}&{\rm WOH} \cite{woh}\\

0.23205^{+0.00004}_{-0.00014}&{\rm ZFITTER} \cite{zfi}\\
\end{array}
\right.
\ee
%
The quoted errors are due to the various options offered by each
library for the treatment of higher-orders in the perturbative
expansion. Since each code has adopted a different renormalization
scheme, a reasonable estimate of the error induced by the scheme
dependence can be obtained by comparing the central values for
$\sin^2\theta^l_{eff}$:
\be
(\delta\sin^2\theta^l_{eff})_{\rm scheme}=
\frac{0.23205-0.23194}{2}=0.00006
\ee
On the other hand, unknown higher-order effects may be sized
by comparing the absolute maximum and minimum
among the values quoted in the previous table.
\be
(\delta\sin^2\theta^l_{eff})_{\rm higher-orders}=
\frac{0.23209-0.23187}{2}=0.00011
\ee
This last uncertainty, which shifts to about 0.00015 if one moves
away from the reference point, is not far from the future experimental
goal.

More that 10 observables have been analyzed in this way in ref. \cite{rep}.
The qualitative conclusion which one can draw is that the intrinsic
theoretical uncertainty on electroweak quantities, mainly due to
unknown two-loop terms in the electroweak expansion, is typically
small if compared with the present experimental precision and,
for $\sin^2\theta^l_{eff}$, with the uncertainty induced
by the hadronic contribution to the photon vacuum polarization;
nevertheless the intrinsic uncertainty is close to the ultimate
experimental precision and only
the explicit evaluation of the missing two-loop electroweak contribution
could reduce it adequately.

As a step towards the realization of this ambitious program,
two-loop electroweak corrections of order $O(G_\mu^2 m_t^2 M_Z^2)$
have been recently computed for the $\rho$ parameter \cite{dff}.
These corrections
are called sub-leading, since they are suppressed with respect to
those explicitly given by the $N_c x_t^2 \Delta\rho^{(2)}$ in eq.
(\ref{drho}) by a factor $\dd\frac{M_Z^2}{m_t^2}$.

The leading
correction $N_c x_t^2 \Delta\rho^{(2)}$ has the property of remaining
unchanged in the limit of vanishing gauge coupling constants $g$ and $g'$
of the SM. This fact suggested the possibility of evaluating it in
the framework of a Yukawa theory, obtained
from the SM by turning the gauge interactions off. Such a theory
describes the ideal case of top mass much larger than the vector boson
and light fermion masses, with gauge interactions negligible compared to
those associated to the top-scalar sector.
The explicit computation of $\Delta\rho$ within the Yukawa limit
of the SM is made possible by the existence
of Ward identities which, in renormalizable gauges, relate vector
boson Green functions to Green functions of corresponding unphysical scalars.
Indeed, amplitudes for unphysical would-be Goldstone bosons can be defined and
worked out in the Yukawa theory. There are various advantages in using
this method. There are less diagrams to compute than in the full theory,
they are more convergent and all the complications related to gauge
theories, such as gauge fixing, ghosts and so on, disappear.
This technique has been successfully
used in the last years to evaluate corrections of $O(G_\mu^2 m_t^4)$ to
the $\rho$ parameter and to the width $\Gamma_b$ \cite{bar}, as well as
those of $O(G_\mu \alpha_s m_t^2)$ to $\Gamma_b$ \cite{deg}.

Contrary to the case of the leading corrections, the evaluation of the
subleading ones requires the use of the full $SU(2)_L\otimes U(1)_Y$
gauge theory. Since they involve the computation of vertex and box
corrections they are not universal, but process dependent. One of the
simplest case which can be considered is the ratio $\rho$
among neutral and charged current
amplitudes in neutrino-lepton scattering at zero momentum
transfer \cite{dff}.
The result can be cast into the following form:
\bea
\rho&=&1+\delta\rho^{(1)}+N_c x_t \delta\rho^{(1)}+\delta\rho^{(2)}\nn\\
&\simeq&\frac{1}{(1-\delta\rho^{(1)~f})}~
(1+\delta\rho^{(1)~b}+\delta\rho^{(2)})
\eea
where $\delta\rho^{(1)}=\delta\rho^{(1)~f}+\delta\rho^{(1)~b}$
is the one-loop result, sum of the separate fermionic and bosonic
contributions ($\delta\rho^{(1)~f}=N_c x_t$).
The two-loop contribution $\delta\rho^{(2)}$, expressed in units of
$N_c x_t^2$, is given by $\Delta\rho^{(2)}$ of eq. (\ref{drho}) plus
a subleading term. These terms are explicitly plotted in fig. 1, as a
function of $m_t$ for three different values of $m_H$.

We see that the subleading corrections are as large as the leading ones.
Actually, for small values of $m_H$, they are even larger, as it could be
expected, due to the accidental smallness of the leading result
for a vanishing $m_H$ \cite{van}. They have the same sign as the leading
ones. For $m_H=250\gev$ and $m_t=176\gev$, the two-loop leading
contribution is about $-3\cdot 10^{-4}$, whereas the sum of leading
plus subleading terms gives about $-6\cdot 10^{-4}$
\footnote{This result is obtained by expressing the fermionic one-loop
contribution in terms of the Fermi
constant $G_\mu$, and the bosonic one-loop part as a function of
${\overline {\rm MS}}$ parameters $\hat\alpha$ and $\hat s^2=
\sin^2\hat\theta_W(M_Z)$. By a different choice of renormalized parameters,
e.g. on-shell ones, the new two-loop correction has approximately the same
numerical value.}.
Finally,
the comparable magnitude of leading and subleading terms raises
doubts about the possibility of determining the full two-loop
contribution from the first few terms of the expansion in $M_Z^2/m_t^2$.

These results, referring to a specific process at zero momentum transfer,
cannot be directly used for LEP observables.
It is however tempting to make a naive extrapolation from $q^2=0$
to the $q^2=M_Z^2$. By using the relations:
\be
\frac{\delta s^2}{s^2}=-\frac{s^2 c^2}{(c^2-s^2)}\delta\rho,~~~~~~~~~~
\frac{\delta M_W}{M_W}=\frac{c^2}{2~(c^2-s^2)}\delta\rho
\ee
where $s$ and $c$ denote the sine and cosine of $\theta^l_{eff}$,
one would obtain, for $m_H=250\gev$ and $m_t=176\gev$
and by including leading plus subleading terms,
the following estimate of the two-loop electroweak corrections on
$\sin^2\theta^l_{eff}$ and $M_W$:
\be
\delta_{EW}^{(2)}(\sin^2\theta^l_{eff})=+0.0002,~~~~~~~~~~
\delta_{EW}^{(2)}(M_W)=-35\mev
\ee

In conclusion, the present level of intrinsic theoretical precision
is sufficient to discuss the existing data. Future experimental
precisions will however require further improvement in the evaluation of
the hadronic contribution $\Delta\alpha_h$ and the inclusion of
two-loop electroweak corrections in the existing libraries.

\vskip 0.5cm
{\bf Acknowledgement}:
We would like to thank D. Bardin for having kindly provided to us a
preliminary version of ref. \cite{rep}.

\vskip 1.0cm
\noindent
{\Large\bf Figure Captions}:

\noindent
Fig. 1: $\delta\rho^{(2)}$ for $\nu_\mu\, e$ scattering, in units $N_c x_t^2$
as a function of $m_t$ for few values
of $m_H$: including only the $O(G_\mu^2 m_t^4)$ contribution $(y)$,
and with both the $O(G_\mu^2 m_t^4)$ and $O(G_\mu^2 m_t^2 M_Z^2)$ terms $(g)$.

\vskip 1.0cm


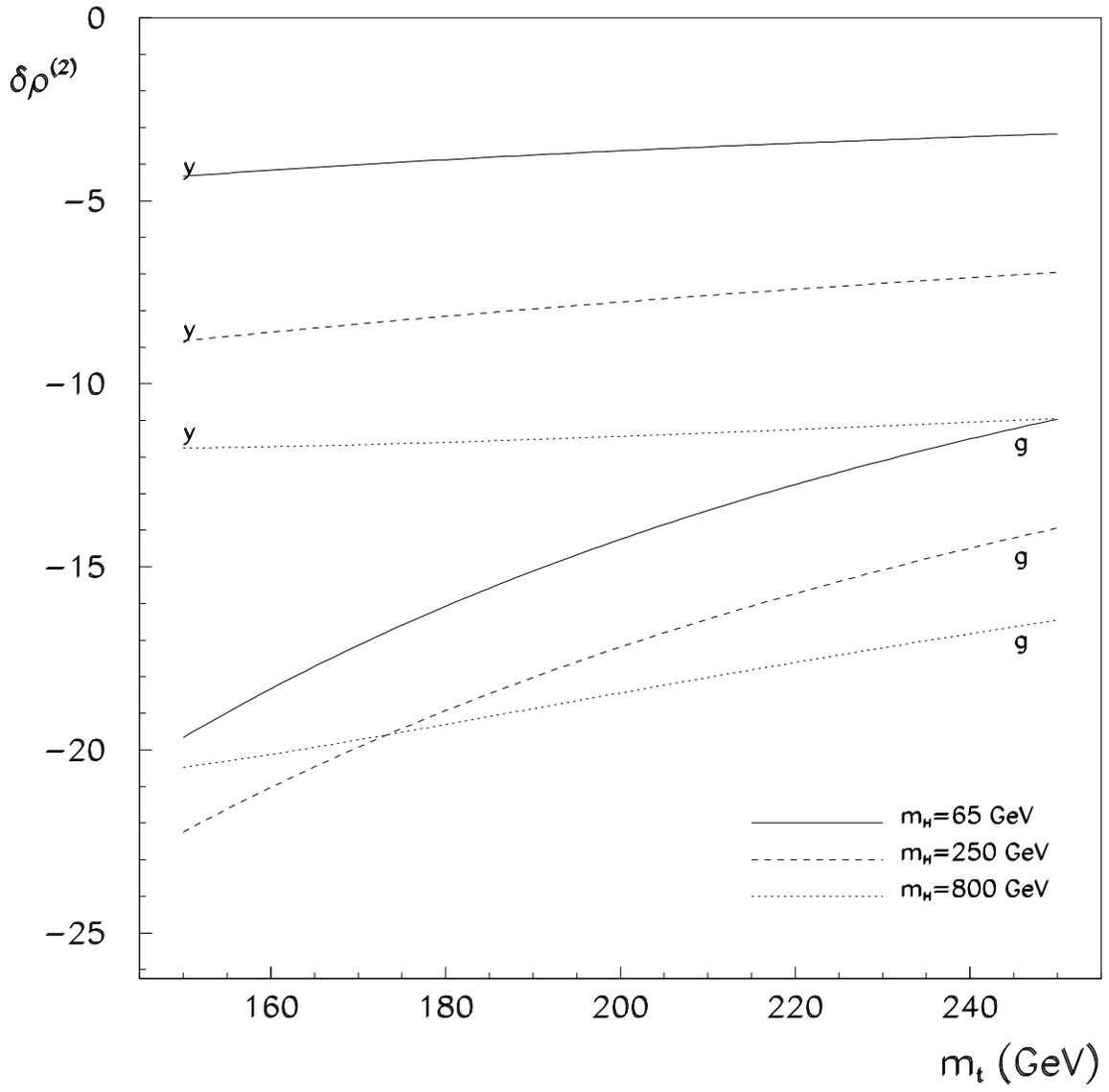
\end{document}